\documentclass[preprint,amsmath,amssymb,showpacs,superscriptaddress]{revtex4}

\usepackage{graphicx}
\usepackage{natbib}
\usepackage[usenames]{color}

\newcommand{\be}{\begin{equation}}
\newcommand{\ee}{\end{equation}}

\def\bear{\begin{eqnarray}}
\def\eear{\end{eqnarray}}

\begin{document}
\title{Curvature in graphene nanoribbons generates \\temporally and spatially focused electric currents}

\author{C. G. Rocha}
\affiliation{Nanoscience Center, Department of Physics, University of Jyv\"askyl\"a, 40014 Jyv\"askyl\"a, Finland} \affiliation{School
of Physics, Trinity College Dublin, Dublin 2, Dublin, Ireland}

\author{R. Tuovinen}
\affiliation{Nanoscience Center, Department of Physics, University of Jyv\"askyl\"a, 40014 Jyv\"askyl\"a, Finland}

\author{R. van Leeuwen}
\affiliation{Nanoscience Center, Department of Physics, University of Jyv\"askyl\"a, 40014 Jyv\"askyl\"a, Finland}

\author{P. Koskinen}
\email[]{pekka.koskinen@iki.fi}
\affiliation{Nanoscience Center, Department of Physics, University of Jyv\"askyl\"a, 40014 Jyv\"askyl\"a, Finland}

\date{\today}
\keywords{Graphene, nanoribbons, electronic structure, transport
properties, wavelet, time-dependent}

\begin{abstract}
Today graphene nanoribbons and other graphene-based nanostructures can be synthesized with atomic precision. But while investigations have concentrated on straight graphene ribbons of fixed crystal orientation, ribbons with intrinsic curvature have remained mainly unexplored. Here, we investigate electronic transport in intrinsically curved graphene nanoribbons coupled to straight leads, using two computational approaches. Stationary approach shows that while the straight leads govern the conductance gap, the presence of curvature blurs the gap and reduces on-off ratio. An advanced time-dependent approach shows that behind the fa\c{c}ade of calm stationary transport the currents run violently: curvature triggers temporally and spatially focused electric currents, to the extent that for short durations single carbon-carbon bonds carry currents far exceeding the steady-state currents in the entire ribbons. Recognizing this focusing is pivotal for a robust design of graphene sensors and circuitries.
\end{abstract}

\maketitle

\section{Introduction}
Graphene is a nanomaterial that can be tailored extensively for device fabrication purposes. It has already been used in field-effect transistors\cite{georgiou2013}, non-volatile memory elements\cite{bertolazzi2013}, logical switches\cite{standley2008}, among other electronic components\cite{geim2007}. A central part in the making of graphene leads, components, and circuitries is the patterning. Today experiments can routinely fabricate graphene nanoribbons (GNRs) in a variety of shapes, widths, and curvatures---even with atomic precision.\cite{cai2010, palma2011} For this reason especially straight GNRs have been investigated much both theoretically and experimentally.\cite{barone_nlett_2006,son_nature_2006,li2008,koskinen_PRL_2008,cai2010,wang_nnano_2011} However, in addition to the straight GNRs, patterned graphene nanoribbons can also contain intrinsic curvature.

Curvature has been demonstrated as semicircular graphene ribbons grown on a templated silicon carbide substrate\cite{sprinkle2010}, as kinked and folded graphene with well-defined kink angles\cite{li2008}, and as nanoribbons sculptured by meniscus-mask lithography with customized edge curvatures\cite{abramova2013}. In addition, curvilinear graphene microcircuits or stamps have been imprinted directly on graphene oxide films\cite{zhang2010}. But even though experiments have frequently realized curved structures, their transport properties have been remained unexplored. The stationary conductance simulations by Wurm \emph{et al.}\cite{wurm2009}, Yin \emph{et al.}\cite{yin2010}, and Qiu \emph{et al.}\cite{qiu2013,qiu2014} are among the few investigations on curvature effects. We think that the additional degrees of freedom created by curvature deserve far more attention than previously invested. Extending the views beyond straight graphene ribbons would help to improve the design of devices and nanocircuits. 

In this work we investigate electronic transport through curved graphene nanoribbons (CGNRs) using two theoretical approaches. First, we model CGNRs using density-functional tight-binding and calculate their transport properties by the regular Landauer approach\cite{nardelli}. These calculations are performed for a systematic set of CGNRs with different ribbon widths, edge curvatures, and lead couplings. Second, with the same set of CGNRs, we simulate transport using an advanced dynamic approach\cite{tuovinen2013, tuovinen2014}, a recent generalization of the Landauer-B\"uttiker formula for time-dependent transport\cite{lb}. This approach provides a transparent analysis of the spatial and temporal dependence of electric currents, including their transient dynamics. The analysis reveals that the stationary approach overlooks the complex processes that occur during hundreds of femtoseconds after switching on the bias voltage: in the time-dependent picture the currents in CGNRs show strong focusing both spatially and temporally.

\section{Curved graphene nanoribbons in transport calculations}

To create models for the CGNRs, graphene disks were first cut out of pristine graphene by setting the origin at a hollow site, removing atoms below a minimum radius $R_i$ and above a maximum radius $R_o$, and then removing the resulting singly-coordinated edge atoms. These disks were then bisected to yield $60^\circ$, $90^\circ$, or $180^\circ$ arcs, to constitute the central scattering part for the transport calculation. Finally, to create the lead electrodes, the arcs were coupled to straight, semi-infinite graphene nanoribbons. The leads were of armchair type (AGNR) or zigzag type (ZGNR), depending on the bisection angle\cite{Wakabayashi2010}. The $60^\circ$ and $180^\circ$ arcs were coupled to two AGNRs and the $90^\circ$ arcs to AGNR on one side and to ZGNR on the other side.

\begin{figure}[t]
\centering
\includegraphics[width=0.4\textwidth]{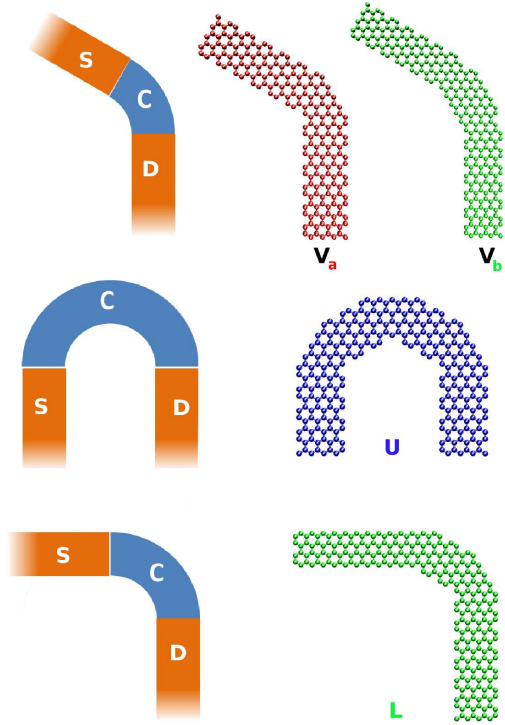}
\caption{(Color online) Curved graphene nanoribbons in transport calculations. Left panels: Schematics of curved ribbons attached to straight ribbons of the same width, forming a typical two-terminal transport device: source (S) and drain (D) electrodes connected via a central conducting device (C). Right panels: Atomic structures of the samples V$_{\text{a}}$, V$_{\text{b}}$, U, and L. Hydrogen atoms are omitted from the figure for clarity.}
\label{fig:structures}
\end{figure}

The three-part systems were saturated by hydrogen to remove the dangling bonds. Note that the systems have no in-plane stresses. We follow the convention to call the three parts the source electrode (S), the central conducting device (C), and the drain electrode (D) (Fig.~\ref{fig:structures}).


We focus our discussion on a representative set of four CGNR samples (Fig.~\ref{fig:structures}). The samples include two $60^\circ$ arcs of different curvature (samples V$_{\text{a}}$ and V$_{\text{b}}$, Fig.~\ref{fig:structures}a). They both have 8-AGNR legs but in V$_{\text{a}}$ the curved parts are shorter and only the outer edge contains zigzag section; in V$_{\text{b}}$ both inner and outer edges contain zigzag sections. The sample U contains a $180^\circ$ arc and the sample L a $90^\circ$ arc (Figs.~\ref{fig:structures}b and c). These four samples are representative to illustrate the effects in transport properties caused by curvature. However, while the discussion focuses on these four samples, the general results are supported by calculations and detailed characterizations of a systematic set of different CGNRs, as presented in Supplementary Information (SI).\cite{SI} 

\section{Computational methods for stationary and time-dependent transport}
\label{sec:methods}
To model the stationary transport, we used density-functional tight-binding (DFTB) method\cite{dftb}. DFTB has been shown to capture the essential features of graphene-based materials well\cite{porezag_PRB_1995,elstner_PRB_1998,kudin_PRB_2001,frauenheim_PSS_2000,koskinen_CMS_2009,kit_PRB_2012}. To prepare the samples, CGNRs were coupled to semi-infinite leads (lengths $\sim 88$~\AA~were sufficient), all edges were hydrogen-passivated to avoid finite-size effects, and the systems were relaxed. Self-consistent DFTB calculations then gave the $\Gamma$-point Hamiltonians for the curved parts, two principal layer units of source and drain electrodes, and their coupling matrices. These Hamiltonians were then used to construct the Green's functions and self-energies for the three parts (L, C, and R) in order to calculate the stationary conductance by the Landauer formula.\cite{nardelli} 

To model the time-dependent transport, we used the advanced approach of Refs.~\cite{tuovinen2013,tuovinen2014}. In this approach Kadanoff-Baym equations\cite{kbaym} are solved analytically to yield a time-dependent generalization of the Landauer-B\"uttiker formula\cite{lb}. Just like the stationary approach, also this approach assumes noninteracting electrons in a system composed of a central scattering part coupled to leads that are described within wide-band limit. Initially all the system parts were in thermal equilibrium at a certain chemical potential $\mu$. At a given time the bias voltage $V_{\text{SD}}=V_{\text{S}}-V_{\text{D}}$ was switched on symmetrically around $\mu$. After the switch-on the reduced one-particle density matrix $\rho(t)$ was evaluated directly from the closed-form expression with a time parameter $t$.\cite{tuovinen2014} Remarkably, the solution for $\rho(t)$ appears in closed form and \emph{requires no time propagation}.

This approach provides a powerful tool for a transparent analysis of electron dynamics. Expressing the density matrix in a \emph{localized} basis (of the system's Hamiltonian) the diagonal elements yield local charge densities in individual atoms, whereas the off-diagonal elements yield currents in individual bonds. The approach thus provides a direct access to spatial and temporal information of electron dynamics.

While the stationary transport was calculated using the full multi-orbital DFTB Hamiltonian, the time-dependent transport was calculated by a single $\pi$-orbital tight-binding approximation with the hopping parameter $\gamma_{\text{cc}}=-2{.}7$~eV. Originally we began by using the full DFTB Hamiltonian also with the time-dependent approach, but it only added complexity to the analysis by giving noisy and complex current profiles within the transient regime. The $\pi$-orbital approximation was chosen because it simplified the analysis but still captured the central features of the time-dependent dynamics, even quantitatively\cite{harju2010,harju2011} (See also Sec.~\ref{sec:1sttrans}.)

\section{Transport in stationary picture}

\begin{figure}[t]
\centering
\includegraphics[width=0.45\textwidth]{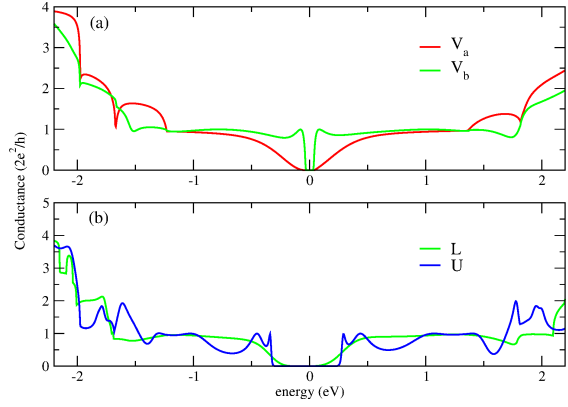}
\caption{ (Color online) Stationary conductance of the four CGNRs of Fig.~\ref{fig:structures}. 
\label{fig:stationary}}
\end{figure}

In Fig.~\ref{fig:stationary} we show the conductance of the four CGNR samples as a function of energy. The transport character of CGNR is dominated by the semiconductor or metallic nature of the AGNR or ZGNR leads, only with a slightly modified transport gap. Still, there are significant differences in the stationary transport of curved and straight ribbons, when compared directly (SI).\cite{SI} Differences result from a pronounced electronic scattering that takes place at the interfaces between armchair and zigzag sections.\cite{wurm2009} 


The transport gap increases when the angle between the leads is increased. The gap is smallest for samples V$_{\text{a}}$ and V$_{\text{b}}$ because the leads are symmetric and the central part creates relatively little scattering. In samples L and U the current has to turn a larger angle, which results in more scattering in the curved part and a larger transport gap.

In straight ribbons the conduction channels open and close abruptly\cite{SI}, but in CGNRs the channel energies are broadened by the asymmetry between outer and inner edges. Thus, one effect of curvature is to blur the transport gap. The blurring can be understood in a simple picture: an incident electron wavefunction propagating in the lead at a certain energy deflects in response to the curvature and creates an energy broadening. This broadening depends on the atomic details of the curved part.

\begin{figure}
\centering
\includegraphics[width=0.6\textwidth]{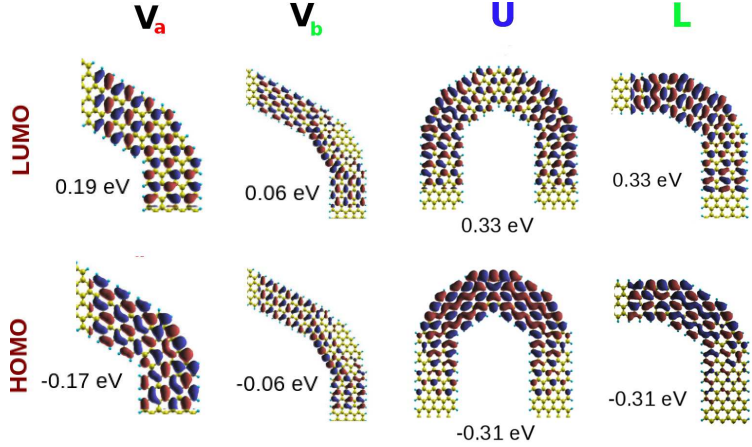}
\caption{(Color online) Electron wave functions for LUMO (lowest unoccupied molecular orbital) and HOMO (highest occupied molecular orbital) in the CGNRs of Fig.~\ref{fig:structures}. 
\label{fig:wfs}}
\end{figure}

The stationary conductance can be understood via electron wavefunction visualizations at the $\Gamma$ point. The highest occupied (HOMO) and lowest unoccupied (LUMO) molecular orbitals reveal both localized and delocalized characteristics (Fig.~\ref{fig:wfs}). States relevant for transport show wavefunctions delocalized across the entire system. However, sample V$_{\text{b}}$ contains zigzag sections at both inner and outer edges and shows strong HOMO and LUMO localizations at the inner edge. The electrons in short zigzag segments are trapped by the neighboring semiconducting AGNR leads, leading to donor-like states near conduction bands and acceptor-like states near valence bands. Moreover, larger angle between the leads creates stronger localization of the frontier orbitals in the curved section, which is part of the origin for the observed large transport gap in L and U samples.

\section{Transport in time-dependent picture}

\subsection{Transient behavior of currents in curved ribbons}\label{sec:1sttrans}


The stationary results suggest that the currents may run less smoothly through curved ribbons than through straight ribbons. To get a more complete microscopic picture of the current dynamics, we used the advanced time-dependent Landauer-B\"uttiker formula, described in Section \ref{sec:methods}. This time-dependent picture does not reduce full transport characteristics to a mere number, the energy-dependent conductance, but gives a more transparent access to both temporal and spatial dependence of the current.  

\begin{figure}[t]
\centering
\includegraphics[width=0.55\textwidth]{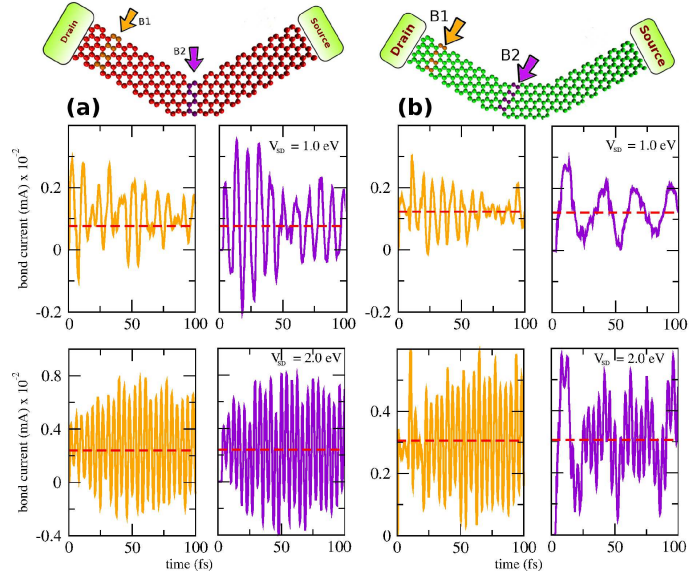}
\caption{(color online) Time-dependent currents in CGNRs. (a) Top figure shows the atomic structure of sample V$_{\text{a}}$ with bridges B1 and B2. The four panels show time-dependent currents through B1 (left) and B2 (right) at $V_{\text{SD}} = 1{.}0$~eV (middle) and at $V_{\text{SD}} = 2{.}0$~eV (bottom). (b) Same as panels a for sample V$_{\text{b}}$. Dashed lines mark the steady-state currents.
\label{fig:td-currents}}
\end{figure}
\begin{figure}[t]
\centering
\includegraphics[width=0.5\textwidth]{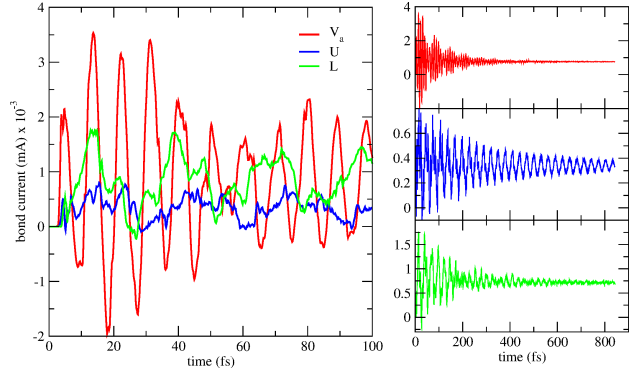}
\caption{(color online) Left: Short-time behavior of currents through samples V$_{\text{a}}$, U, L at $V_{\text{SD}} = 1{.}0$ eV. Right: The corresponding long-time behaviors, showing current saturation. Currents were calculated through a bridge in the middle of the curved section. 
\label{fig:bcvul}}
\end{figure}

In the time-dependent approach, the bias voltage $V_{\text{SD}}$ was switched on at $t=0$ and the transient dynamics was calculated until a steady state was reached. The bias voltages ranged from $V_{\text{SD}}=0{.}2$~eV to $2{.}0$~eV. The coupling strengths were chosen so that the dissipation rate to the semi-infinite leads was $\gamma=0{.}1$ eV, for both source and drain. For our samples these conditions required saturation times up to $1$~ns (see Fig.~\ref{fig:bcvul} and Ref.~\cite{SI}). The transient dynamics were first analyzed as currents through two bridges, B1 and B2, which were calculated as the sum over individual bond-currents (Fig.~\ref{fig:td-currents}).

From the moment the bias is switched on, the current at B1 grows rapidly and starts to oscillate over hundreds of femtoseconds until the oscillations damp towards a steady state. To reach B2 the wavecrest needs to travel $\sim 30$~ \AA, which is seen as a short delay in the bridge current. This delay implies current velocity of $\sim 10$~\AA/fs, which equals the Fermi-velocity in graphene, as expected.\cite{graphene_rev} The transients at both bridges are characterized by slow oscillations superimposed by fast oscillations associated with multiple intra-ribbon and ribbon-lead state transitions. The slowest oscillations originate from lead-to-lead reflections created by the charge density wave after switching on the bias. These slow oscillations are visible in the long-time plots of bridge currents (Fig.~\ref{fig:bcvul} and SI\cite{SI}). High bias reveals another type of slow oscillation that originates from multiple scatterings of the charge density waves trapped in the curved part.

\begin{figure}[t]
\centering
\includegraphics[width=0.55\textwidth]{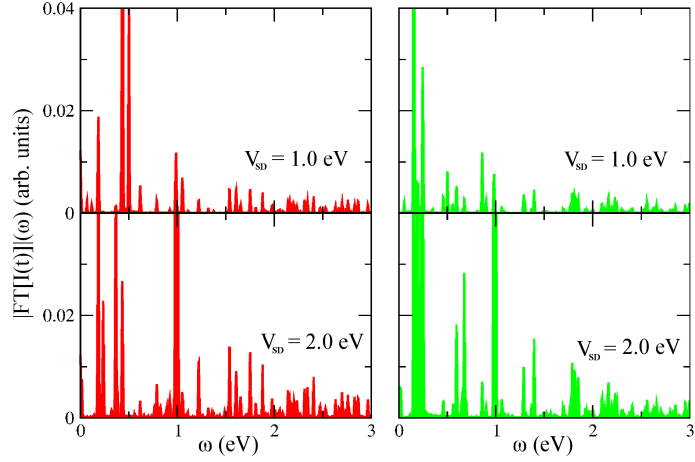}
\caption{(color online) Fourier transforms of time-dependent B2 currents of Fig.~\ref{fig:td-currents}. Samples are V$_{\text{a}}$ (left panels) and V$_{\text{b}}$ (right panels) at the shown biases. 
\label{fig:fourier}}
\end{figure}

In straight GNRs the currents are regular and flow in the same direction\cite{tuovinen2014}, but in CGNRs they are more complex. The curved parts act as strong scattering centers for the electronic current and cause direction reversal, especially at low bias. Such backward currents are evident when looking at animations of bond current (see SI\cite{SI}). The dynamics of backward currents in CGNRs are governed by two length scales: lead-to-lead and lead-to-curved part distances. These distances create rich interference patterns upon multiple scatterings at the interfaces. The interference results in dense frequency spectra of the Fourier-transformed (FT) currents (Fig.~\ref{fig:fourier}).

\begin{figure}[t]
\centering
\includegraphics[width=0.55\textwidth]{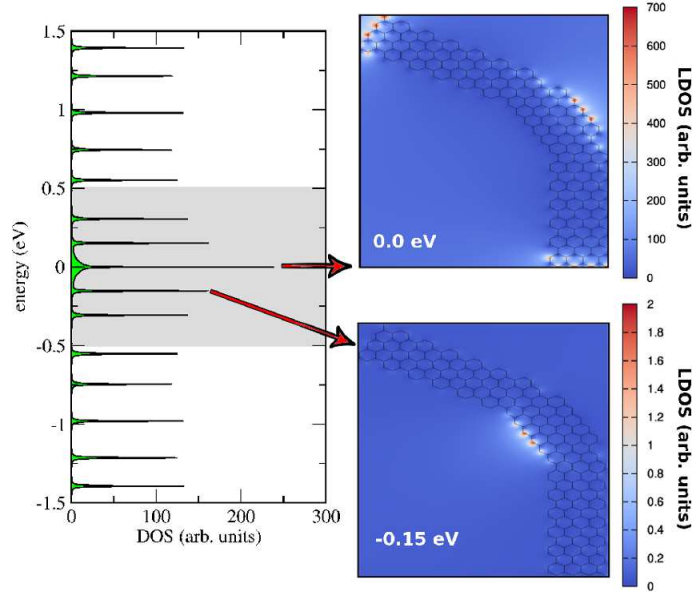}
\caption{(color online) Density of states (DOS) of sample V$_{\text{b}}$. Shaded region shows the $1{.}0$~eV bias window. Right insets show the local density of states (colormaps) for the selected DOS peaks at $0$ eV and $-0{.}15$ eV.
\label{fig:new_spectral}}
\end{figure}

It is instructive to investigate some transitions more closely. For instance, the time-dependent currents of V$_{\text{b}}$ at $V_{\text{SD}}=1.0$~eV exhibit well-defined oscillations at $30$ fs period, and the FT spectra shows an intense peak at the corresponding energy $\omega \approx 0.15$~eV. This transition can be identified by looking at the local density of states (LDOS) at energies suggested by the spectral function. The local density of states at energies $0$~eV and $-0.15$~eV show elongated edge states at inner and outer edges of the curved part (Fig.~\ref{fig:new_spectral}). Thus, the observed electronic excitations at $\omega \approx 0.15$ eV corresponds to edge-state transitions taking place inside the CGNR. This oscillation becomes masked if $V_{\text{SD}}$ is increased because additional electronic transitions start taking place within the enlarged bias window. 

Here we note that the single $\pi$-orbital model compares well with the full DFTB model. Both models yield frontier orbitals of similar edge-localized nature and similar level structure (Figs.~\ref{fig:wfs} and \ref{fig:new_spectral}), even if quantitative differences exist. Our central results are not affected by the choice of the model. 

In Fig.~\ref{fig:bcvul} we compare the transient currents in samples V$_{\text{a}}$, L and U. Currents are calculated at $V_{\text{SD}} = 1{.}0$~eV through bridges in the middle of the curved parts. As a result, smaller angle CGNRs yield larger oscillation amplitudes in the transient regime. The sample V$_{\text{a}}$ with the shortest arc reaches the steady-state current fast, within $\sim 300$ fs. This is because the curved part is relatively short, with a zigzag section only on the outer edge, which means a strong coupling between the lead electrodes and a short equilibration time. The sample U, on the contrary, has a curved part that is longer and more structured, which means weaker coupling between the lead electrodes: current scatters more, creating smaller current fluctuations due to interference between scattered currents, and ultimately a longer equilibrium time. Sample L, with the same reasoning, has an equilibration time intermediate between samples V$_{\text{a}}$ and U.

\begin{figure}[h]
\centering
\includegraphics[width=0.4\textwidth]{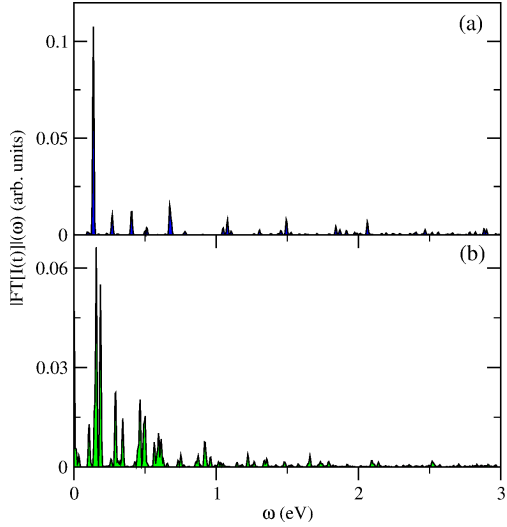}
\caption{Fourier transforms of the time-dependent currents in Fig.~\ref{fig:bcvul} for (a) sample U and (b) sample L at $V_{\text{SD}} = 1{.}0$ eV. 
\label{fig:fourierUL}}
\end{figure}

The spectral analysis for sample U shows an intense transition at $\omega \approx 0.14$~eV, followed by several weaker transitions at higher energies (Fig.~\ref{fig:fourierUL}). Sample L shows richer Fourier spectrum. Consider, in particular, the triplet of intense transitions at $0.10$ eV, $0.16$ eV, and $0.18$ eV. These transitions can be tracked down to the triplet-like edge states at $0$~eV that originate from three zigzag-sections in the structure (see Fig.~\ref{fig:structures} and spectral function in SI \cite{SI}). Therefore, this triplet of excitations, as well as the other triplets at higher energies, correspond to transitions between the edge states at $0$~eV and states at other energies.

The above analysis shows how FT provides some insight to the transitions during current transients. However, it misses all the temporal information. FT is unable to provide either the times or the durations for the activities of certain transitions. FT peak intensities give sort of averaged-out information about how long durations certain frequencies are present; a more accurate temporal analysis however requires an alternative analysis tool.

\subsection{Temporal and spatial characterization:\\ Wavelet analysis}


The alternative analysis tool that gives better insight into the complex back-and-forth currents in CGNR is Wavelet decomposition\cite{wavelet_book,hp,wavelet_friendly}. In this method the time-domain signal is convolved by an element of a set of basis functions called wavelets. For our current results we tested three types of wavelets: Haar\cite{haar}, Morlet\cite{jordan1997,baker2012}, and Ricker\cite{ryan1994,mexh}, the last one being the second derivative of a Gaussian function. 
We will show only results using Ricker wavelet, as it provided the cleanest analysis. 

The continuous wavelet transform (CWT) of the current $I(t)$ convolved by a given wavelet $\psi$ is defined as
\begin{equation*}
C(s,\tau) = \frac{1}{\sqrt{s}}\int_{-\infty}^\infty I(t)\, \psi^*\left ( \frac{t-\tau}{s} \right ) dt,
\end{equation*}
\noindent where $\psi^*$ is complex conjugate of the wavelet shifted by $\tau$ and scaled by the dimensionless parameter $s$. This transformation maps the data into $s \times \tau$ space, where $\tau$ is related to time and $s$ can be related to frequencies via $f_s = f_c / (s\Delta)$, where $\Delta$ is the sampling period and $f_c$ is the center frequency of the wavelet. This frequency is calculated by associating a periodic function of frequency $f_c$ with the respective wavelet in such way that the function approximately delineates the wavelet form. The frequency is related to real time directly as $\tau_s = 1/f_s$. With fs as our time unit, we used Ricker wavelets with a center frequency of 0.25 fs$^{-1}$ and sampling of $\Delta = 0.16$ fs.


The advantage of wavelet analysis is that it enables us to distinguish between rapid and slow current fluctuations by choosing the scale $s$. Currents can thus be analyzed simultaneously at different frequencies and at different times. By searching and analyzing intense wavelet coefficients we can identify instants and frequencies for eminent and possibly interesting events. We calculated CWT on the currents by using the Wavelet Toolbox in Matlab\cite{toolbox} and scales large enough to scan all the frequencies relevant to our samples.

\subsection{Spatial character of time-dependent currents}


\begin{figure}
 \centering
\includegraphics[width=0.7\textwidth]{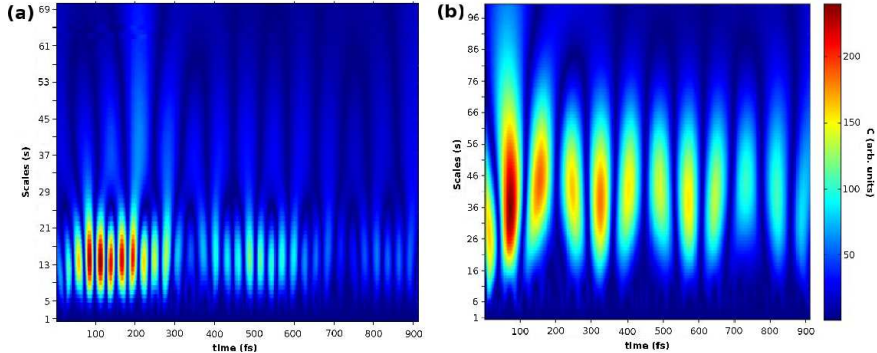}
\caption{Wavelet coefficients (colormap) as a function of scale ($s$) and time ($\tau$) for currents through bridges B2 in (a) sample V$_{\text{a}}$ and (b) in sample V$_{\text{b}}$ at $V_{\text{SD}} = 1.0$~eV.
\label{fig:wavelets1}}
\end{figure}

The wavelet analysis of bridge currents in samples V$_{\text{a}}$ and V$_{\text{b}}$ reveals frequency-dependent periodic structures of intense patterns (Fig.~\ref{fig:wavelets1}). For sample V$_{\text{a}}$ the most intense wavelet components around $s=13$-$15$ correspond to periods of $\tau_s \approx 9.0$ fs and energy of 0.46 eV. This energy agrees quantitatively with  the fourier peak at $\omega \approx 0.45$ eV (Fig.~\ref{fig:fourier}). Wavelet analysis also gives the ``life time'' of this mode: intensity peaks repeat every $\Delta \tau \sim 14-20$~fs and they dominate over other modes until 300 fs after which intensity starts to fade away. The modes with scales $s>35$ are the low-frequency modes with $\omega < 0.065$ eV that correspond to the background oscillations seen in the long run simulations\cite{SI}. Similarly, for sample V$_{\text{b}}$ the most intense wavelet componets around $s=41$ correspond to $\tau_s \approx 27$ fs and energies $\approx 0.15$ eV, in agreement with the spectral analysis (Fig. \ref{fig:new_spectral}). Intensity peaks repeat every $\Delta \tau \sim 55$-$60$~fs and remain pronounced up to $\sim650$~fs. The general current features in V$_{\text{b}}$ are simpler than in V$_{\text{a}}$ and most wavelet componets retain their relative intensities during the course of time. The differences in these features can again be attributed to the longer curved part in V$_{\text{b}}$ that more effectively delays the transmission of the current. The observed timescales provide a valuable information to help optimizing the detection of such transient responses; one could use a probing wave with the same period $\Delta \tau$ to discern the characteristic modes.

\begin{figure}[t]
 \centering
\includegraphics[width=0.45\textwidth]{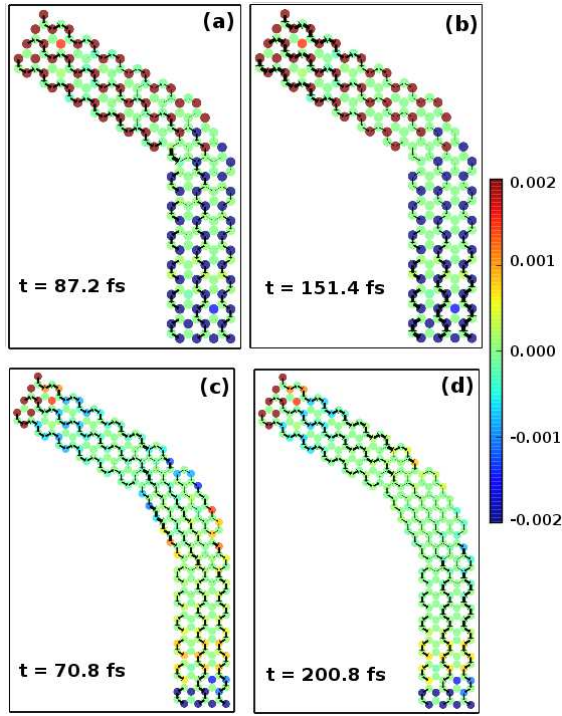}
\caption{ Snapshots of bond currents (black arrows) and charge variations (colormap; units of electron charge) for samples V$_\text{a}$ (a and b) and V$_\text{b}$ (c and d) at $V_{\text{SD}} = 1{.}0$~eV and at the instants shown in the panels.
\label{fig:snaps_V}}
\end{figure}

As discussed above, the wavelet analysis gives the possibility to choose interesting instants to further investigate not only currents through chosen bridges but bond current patterns within the entire samples. Wavelet analysis also enables identifying current ``fingerprints'' for each sample. Let us investigate few current pattern snapshots for samples V$_\text{a}$ and V$_\text{b}$, with instants chosen so that one instant involves an intense wavelet amplitude and the other instant does not (Fig.~\ref{fig:snaps_V}). In both samples currents follow straight paths along the armchair leads and go through the inner zigzag edge in the curved part. Atoms directly connected to source and drain show the largest positive and negative charge density variations, and the spreading of the charge depends on the applied bias, on the length of the curved part, and on the electronic character of the AGNR lead. In sample V$_{\text{a}}$ the charge profile has more internal contrast, with density variations occurring within the entire sample, but with an extremely sharp transition from positive to negative density variations. This is because the leads are metallic 8-AGNRs that enable quick spreading of the charge density waves from the electrodes all the way to the short curved part. In sample V$_{\text{b}}$ the leads are also metallic but the curved part is longer, creating a more gradual transition from positive to negative charge variations and a smoother localization of the current along the inner edge.

\begin{figure}[t]
 \centering
\includegraphics[width=0.45\textwidth]{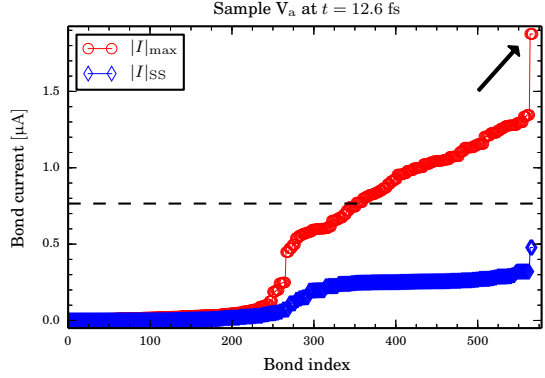}   
\caption{Bond currents in sample V$_{\text{a}}$ at $V_{\text{SD}} = 1{.}0$~eV as a function of bond index (indexing wrt. increasing current). Plotted are the instantaneous bond currents at $t = 12.6$~fs (red circles), the bond currents at the steady-state (blue diamonds), and the overall steady-state current through the sample (dashed line). Arrow points to two singular bond currents that instantaneously carry almost $250$~\% of the overall steady-state current through the sample.
\label{fig:focus}}
\end{figure}

However, the striking feature in the snapshots discussed above is the strong localization or focusing of the current. We analyze the current in the sample V$_{\text{a}}$ by choosing the instant $t=12{.}6$~fs based on the occurrence of the first visible peak in the wavelet amplitude. By creating a simple list of bond currents at that instant, we observe an extreme spatial focusing of the current (Fig.~\ref{fig:focus}); as few as two bonds carry instantaneous currents that are almost $250$~\% of the overall steady-state current through the sample. At this instant also many other bonds carry currents larger than the steady-state current. Another instant with smaller wavelet amplitudes shows much weaker and less inhomogeneous currents. Therefore, the curvature seems to generate electric currents that show high temporal and spatial focusing. This is our central result. Such focusing is absent in straight GNRs, which show temporally and spatially far more homogeneous currents.\cite{tuovinen2014}

\begin{figure}[t]
 \centering
\includegraphics[width=\columnwidth]{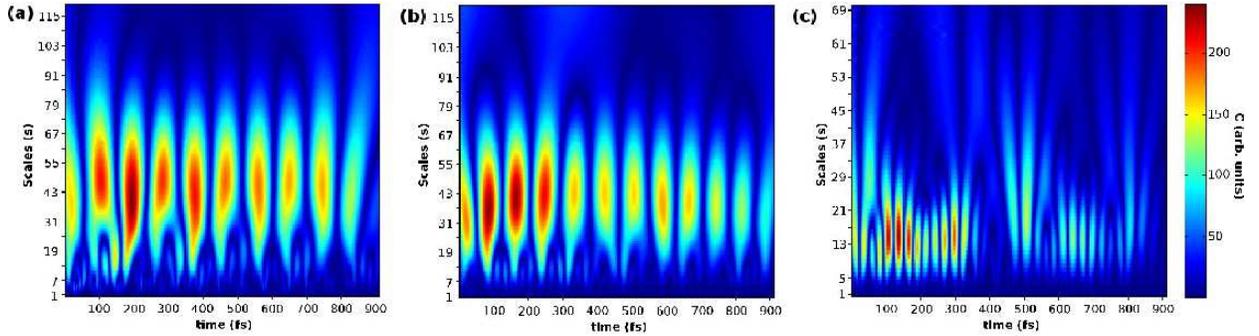}
\caption{Wavelet amplitudes (colormaps) as a function of scale ($s$) and time ($\tau$) for currents through the curved sections in (a) sample U, (b) sample L, and (c) sample V$_{\text{a}}$ with an impurity site, all at $V_{\text{SD}} = 1.0$~eV.
\label{fig:wavelets2}}
\end{figure}

\begin{figure}[t]
\centering
\includegraphics[width=0.55\textwidth]{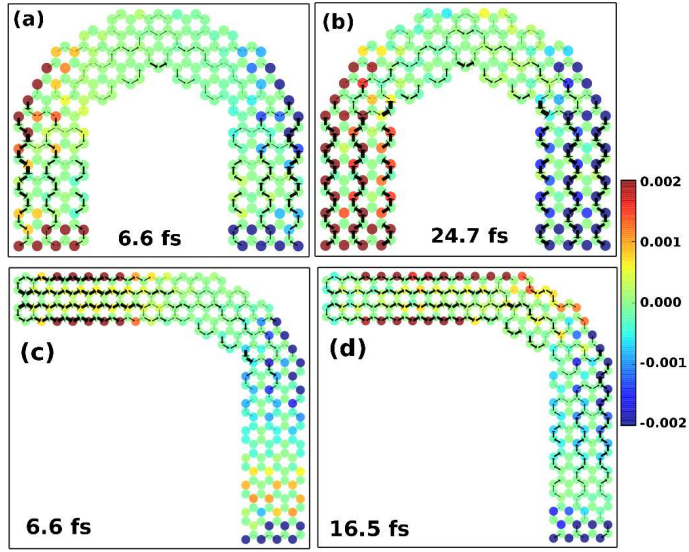}
\caption{Snapshots of the bond currents (black arrows) and the charge variations (colormap; units of electron charge) for samples U (a and b) and samples L (c and d) at $V_{\text{SD}} = 1{.}0$ eV and at times shown in the panels.
\label{fig:snapsUL}}
\end{figure}

The current focusing is a generic feature common to all curved samples. For example, corresponding to the currents in samples U and L (Fig.~\ref{fig:bcvul}), Fig.~\ref{fig:wavelets2} shows the wavelet analysis and Fig.~\ref{fig:snapsUL} selected snapshots of two instants. The wavelet amplitudes and snapshots show current focusing, which is particularly evident in animations (see SI~\cite{SI}). Sample U shows strong currents along the armchair leads and along the inner edges of curved section. Sample L has one lead zigzag and the other lead armchair, and this asymmetry becomes clearly vi\-sib\-le. Charge density variations occur along the sample irregularly, which happens because the zigzag lead supports more current paths than the armchair lead, at least for this type of CGNR. The difference in the propagation speeds in the different leads quickly create a non-uniform and irregular charge variations distributions across the sample.

The wavelet analysis can also provide information about impurities. We repeated the calculations with sample V$_{\text{a}}$, but this time with a model for an impurity atom adsorbed on top of a carbon at the bridge B2. The impurity is modeled by an on-site energy of $-0{.}2$~eV and a modified nearest-neighbor hopping parameter $\gamma_{\text{i}}=0.8 \gamma_\text{cc}=-2{.}16$~eV. This is a crude model for an impurity that causes weaker bonding to its neighboring carbon atoms.\cite{robinson2008} As a result of the impurity, the features of wavelet amplitudes get notably perturbed (compare Fig.~\ref{fig:wavelets1}a with Fig.~\ref{fig:wavelets2}c). The fairly regular wavelet patterns of the pristine sample V$_{\text{a}}$ get disrupted, as becomes evident already within the first $100$~fs. For instance, the second large-intensity region in $s-\tau$ -space shifts towards larger scales, indicating that this particular mode becomes red-shifted. This shifting was confirmed by FT spectrum, which showed a peak at $0.20$~eV for the pristine sample and at $0.18$~eV for the impurity sample. The shifting occurs with scales $s\approx 21$-$39$, corresponding to periods between $13.8$~fs and $25.7$~fs. However, the most important effects of the impurity are the changes in current paths. Animations show how the impurity affects the charge distributions within the entire sample and even alters the current paths from outer to inner edge of the curved section (see SI \cite{SI}). In CGNRs the electric currents can be extremely sensitive to single impurities: if impurities get adsorbed to sites with focused currents, the entire current patterns may have to rearrange. This can drastically change sample conductance.   


\section{Discussion and Conclusions} 

Today experiments, by using sophisticated microscopy techniques, are able to investigate the dynamics of nanoscale materials at nanosecond\cite{wong2013} and at picosecond\cite{pasi2010} time scales, even if not yet quite at femtosecond time scales. However, even if direct measurements at the femtosecond scale should still remain out of reach, the events and dynamics at that scale nevertheless reflect behavior also in longer time scales. Our theoretical approach provides a transparent and intuitive perspective into the femtosecond dynamics and it also scales efficiently with respect to system size; it can be used to study experimentally relevant systems for phenomena with scales ranging from femtosecond dynamics to steady-state properties. In the future we will extend the method for other elements and other types of time-dependent perturbations. 

The simulations using this method show that curvature in graphene nanoribbons causes significant scatterings via impurity-like states, degrading the conducting channels in the vicinity of the transport gap. These scatterings cause strong temporal and spatial focusing, as seen in the visualizations enabled by wavelet analysis. The temporal focusing related to transient phenomena, also observed in other studies and even in straight GNRs\cite{perfetto2010,xie2013,tuovinen2014}, highlights the necessity of time-dependent approach to quantum transport. The curvature-generated spatial focusing is of fundamental importance and could be exploited in various applications, such as in chemical sensing and in the design of graphene circuitries in general.

Scanning transmission electron microscopy (STEM) images of graphene constrictions exposed to adatoms have revealed that their sensitivity, as characterized by susceptible edge-states, can be superior to 'bulky' graphene.\cite{warner2013} Edge-state detectors, such as solid-state nanopore sensors applied to ultrafast DNA decoding, are currently under intense research.\cite{dekker2010,garaj2010,merchant2010}. Decoding with single base resolution is challenging \cite{haque2012}, and one of the related critical challenges is a sufficient spatial and temporal resolution. In this regard curved graphene structures might be suitable systems: currents at their edges are intense and highly focused.


 

\section*{Acknowledgements}
We would like to acknowledge the Academy of Finland for funding (projects 283103, 251216) and the CSC - IT Center for Science in Finland together with the Trinity Centre for High Performance Computing (TCHPC) in Ireland for the computational resources. R.T. wishes to thank V\"ais\"al\"a Foundation of The Finnish Academy of Science and Letters for financial support.


\end{document}